\documentclass{aa}
\usepackage{amsmath,amssymb,epsf}

\makeatletter
\def\thebibliography#1{\section*{REFERENCES}
 \addcontentsline{toc}{section}{REFERENCES}
 \list{}{\labelwidth\z@
         \leftmargin 1.5em
	 \itemsep \z@
	 \itemindent-\leftmargin}
 \small\raggedright
 \parindent\z@
 \parskip\z@ plus .1pt\relax
 \def\newblock{\hskip .11em plus .33em minus .07em}
 \sloppy\clubpenalty4000\widowpenalty4000
 \sfcode`\.=1000\relax
}

\def\@biblabel#1{}
\def\@bcite#1#2{(#1\if@tempswa , #2\fi)}
\def\@pcite#1#2{#1\if@tempswa , #2\fi}
\def\@citefmta#1#2{#1 (#2)}
\def\@citefmtb#1#2{#1 #2}
\let\citefmt=\@citefmta

\def\@citex[#1]#2{\if@filesw\immediate\write\@auxout{\string\citation{#2}}\fi
  \def\@citea{}\@cite{\@for\@citeb:=#2\do
    {\@citea\def\@citea{;\penalty\@m\ }\@ifundefined
    {b@\@citeb}{{\bf ?}\@warning
{Citation `\@citeb' on page \thepage \space undefined}}%
{\csname b@\@citeb\endcsname}}}{#1}}

\def\cite{\@ifnextchar [{\let\citefmt=\@citefmtb
                          \let\@cite=\@bcite\@tempswatrue \@citex}
                        {\let\citefmt=\@citefmtb
                          \let\@cite=\@bcite\@tempswafalse \@citex[]}}
\def\pcite{\@ifnextchar [{\let\citefmt=\@citefmtb
                          \let\@cite=\@pcite\@tempswatrue\@citex}
                        {\let\citefmt=\@citefmtb
                          \let\@cite=\@pcite\@tempswafalse\@citex[]}}
\def\scite{\@ifnextchar [{\let\citefmt=\@citefmta
                          \let\@cite=\@pcite\@tempswatrue\@citex}
                        {\let\citefmt=\@citefmta
                          \let\@cite=\@pcite\@tempswafalse\@citex[]}}
\makeatother

\def\hMpc{\ifmmode{h^{-1}{\rm Mpc}}\else{$h^{-1}$Mpc}\fi}
\def\d{{\rm d}}
\def\bx{{\mathbf{x}}}
\def\CB{{\cal B}}
\def\CX{{\cal X}}
\def\BR{{\mathbb{R}}}
\newcommand{\aanda}{A\&A}
\newcommand{\aas}{A\&AS}

\newcommand{\mnras}{MNRAS}
\newcommand{\nat}{Nat}

\begin{document}

\thesaurus{0.2(12.12.1, 12.03.4, 11.03.1)}

\title{Regularity in the distribution of superclusters?}

\author{Martin Kerscher}

\institute{Sektion Physik, Ludwig--Maximilians Universit\"{a}t,  
Theresienstr. 37, 80333 M\"{u}nchen, Germany\\
email: kerscher@stat.physik.uni-muenchen.de}

\date{Received 7 April 1998 -- accepted 1 May 1998}

\maketitle

\begin{abstract}
We use  a  measure of clustering derived   from the  nearest neighbour
distribution and the void probability  function to distinguish between
regular and clustered structures.  This measure  offers a succinct way
to   incorporate    additional  information  beyond    the  two--point
correlation  function.   Application to   a supercluster  catalogue by
{}\scite{einasto:supercluster_data}    reveals no   clustering  in the
distribution   of  superclusters.   However,    we  show    that  this
supercluster catalogue is severely  affected by  construction effects.
Taking  these biases into  account we still  find some indications for
regularity on the largest scales, but  the significance is only at the
one--$\sigma$ level.
\end{abstract}

\section{Introduction}

In  a  recent paper  {}\scite{einasto:120mpc}  report a   peak in  the
3D--power spectrum of a catalogue of clusters on scales of 120\hMpc.
{}\scite{broadhurst:large-scale} observed  periodicity on  approximately
the same scales in an analysis of 1D--data from a pencil--beam 
redshift survey.
As is well  known from the theory  of fluids, the regular distribution
(e.g.\  of molecules  in  a hard--core  fluid)  reveals  itself  in an
oscillating   two--point correlation function    and   a peak in   the
structure function respectively (see e.g.~\pcite{hansen:theory}).
In accordance with this an oscillating two--point correlation function
$\xi_2(r)$ or at least a first peak was  reported on approximately the
same  scale ({}\pcite{kopylov:possible},  {}\pcite{mo:typical_scales},
{}\pcite{fetisova:features}, and {}\pcite{einasto:supercluster_II}).

In  this  paper    we    analyze  the  supercluster     catalogue   of
{}\scite{einasto:supercluster_data} which     was  constructed from an
earlier    version         of     the      cluster     catalogue    by
{}\scite{andernach:current}    using a friend--of--friends  procedure.
With methods  based  on  the nearest   neighbour distribution and  the
spherical contact  distribution we  can   show that this  supercluster
catalogue is regular with 95\% significance.
However,  taking into account  the selection and construction effects,
the high significance vanishes and we only find  some indication for a
regular distribution on  large scales, showing that  this supercluster
catalogue  is seriously  affected by  the  construction method  with a
friend--of--friends procedure.

This paper is   organized as follows.   In Sect.~\ref{sect:methods} we
discuss our   methods.  Through   some  examples, we    illustrate the
properties of the $J$-function and  show that it  offers a concise way
to incorporate information about correlations of arbitrary order.
The analysis of the supercluster  distribution and of   a set of  mock
supercluster catalogues is presented  in Sect.~\ref{sect:results}.  We
summarize our results in Sect.~\ref{sect:conclusion}.

\section{Methods} \label{sect:methods}

To analyze the set  of points $\CX=\{\bx_i\}_{i=1}^N$, $\bx_i\in\BR^3$
given by  the redshift  coordinates\footnote{Throughout this article we
measure  length in units  of \hMpc, with $H_0   = 100h\ {\rm km}\ {\rm
s}^{-1}\ {\rm Mpc}^{-1}$.}  of the  superclusters we use the spherical
contact distribution $F(r)$, i.e.\ the  {\em distribution function  of
the distances $r$ between an arbitrary point and the nearest object in
$\CX$}.   $F(r)$ is equal to the  expected fraction of volume occupied
by points  which are  not farther away  than  $r$ from the objects  in
$\CX$.  Therefore, $F(r)$ is equal to the  volume density of the first
Minkowski    functional    as    introduced    into   cosmology     by
{}\scite{mecke:robust}.
As another tool we use the nearest neighbour distribution $G(r)$, that
is defined as  the {\em distribution  function of distances  $r$ of an
object in $\CX$ to the nearest other object in $\CX$}.
For a   Poisson distribution the  probability  to  find  a  point only
depends on the  mean number density  $\overline{\rho}$, leading to the
well--known result
\begin{equation} \label{eq:FG_poi}
G_{\rm P}(r) = 1 - \exp\left(- \overline{\rho} \frac{4 \pi}{3} r^3\right) 
= F_{\rm P}(r).
\end{equation}
Recently, {}\scite{vanlieshout:j} suggested to use the ratio
\begin{equation}
J(r) = \frac{1-G(r)}{1-F(r)}
\end{equation}
as a probe for clustering of a point distribution.  
For    a   Poisson   distribution  $J(r)=1$  follows    directly  from
Eq.~(\ref{eq:FG_poi}).   As   shown   by  {}\scite{vanlieshout:j},   a
clustered   point  distribution  implies  $J(r)\le1$,  whereas regular
structures are indicated by  $J(r)\ge1$.  Typical clustered structures
are produced  by Neyman--Scott   processes {}\cite{neyman:statistical}
which have been used  to model the  distribution of galaxies.  Regular
structures are seen  in a periodic,   or a crystalline  arrangement of
points.  In a statistical  sense,  and opposed to  clustering, regular
(``ordered'') structures are also seen in liquids.
Qualitatively one may explain the behaviour of $J(r)$ as follows:
\begin{itemize}
\item
In a clustered distribution of points $G(r)$ increases faster than for
a random  distribution of   points,  since  the nearest  neighbour  is
typically in the   close surroundings.  $F(r)$  increases  more slowly
than for a random distribution, since  an arbitrary point is typically
inbetween the clusters.  These two effects give rise to $J(r) \le 1$.
\item
On the  other  hand,  in a   regular distribution of   points,  $G(r)$
increases more slowly than for a  random distribution of points, since
the nearest neighbour is typically at a finite characteristic distance
(e.g.~in the case of  a crystal).  $F(r)$  increases faster, since the
typical distance from a random point to a point on a regular structure
is smaller. These two effects cause $J(r)$ to be greater than unity.
\item
$J(r) = 1$ indicates the borderline between clustered and regular
structures.
\end{itemize}
%

\subsection{Gaussian approximation}

\begin{figure}
\begin{center}
\epsfxsize=6cm
\begin{minipage}{\epsfxsize} \epsffile{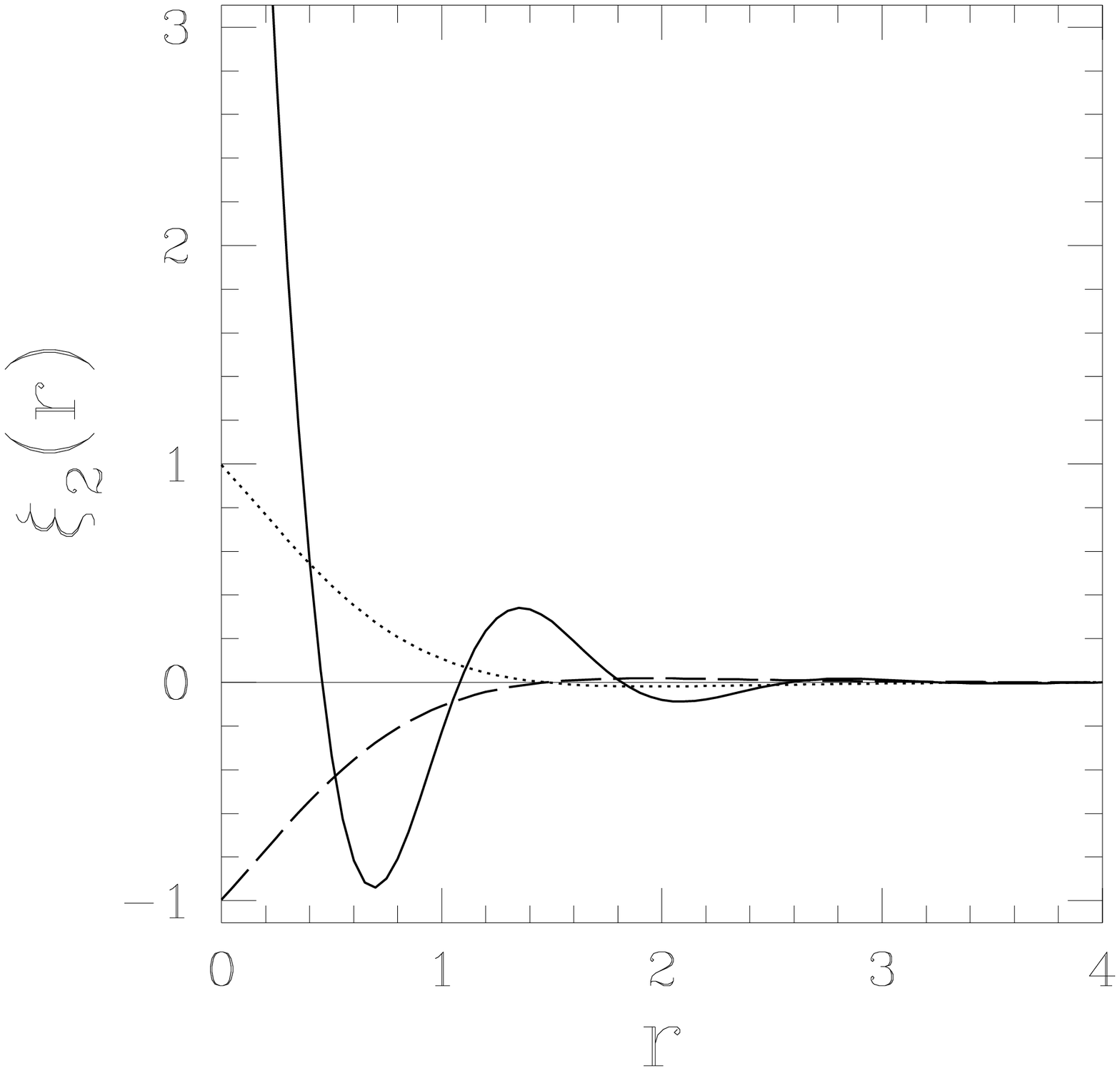} \end{minipage} 
\epsfxsize=6cm
\begin{minipage}{\epsfxsize} \epsffile{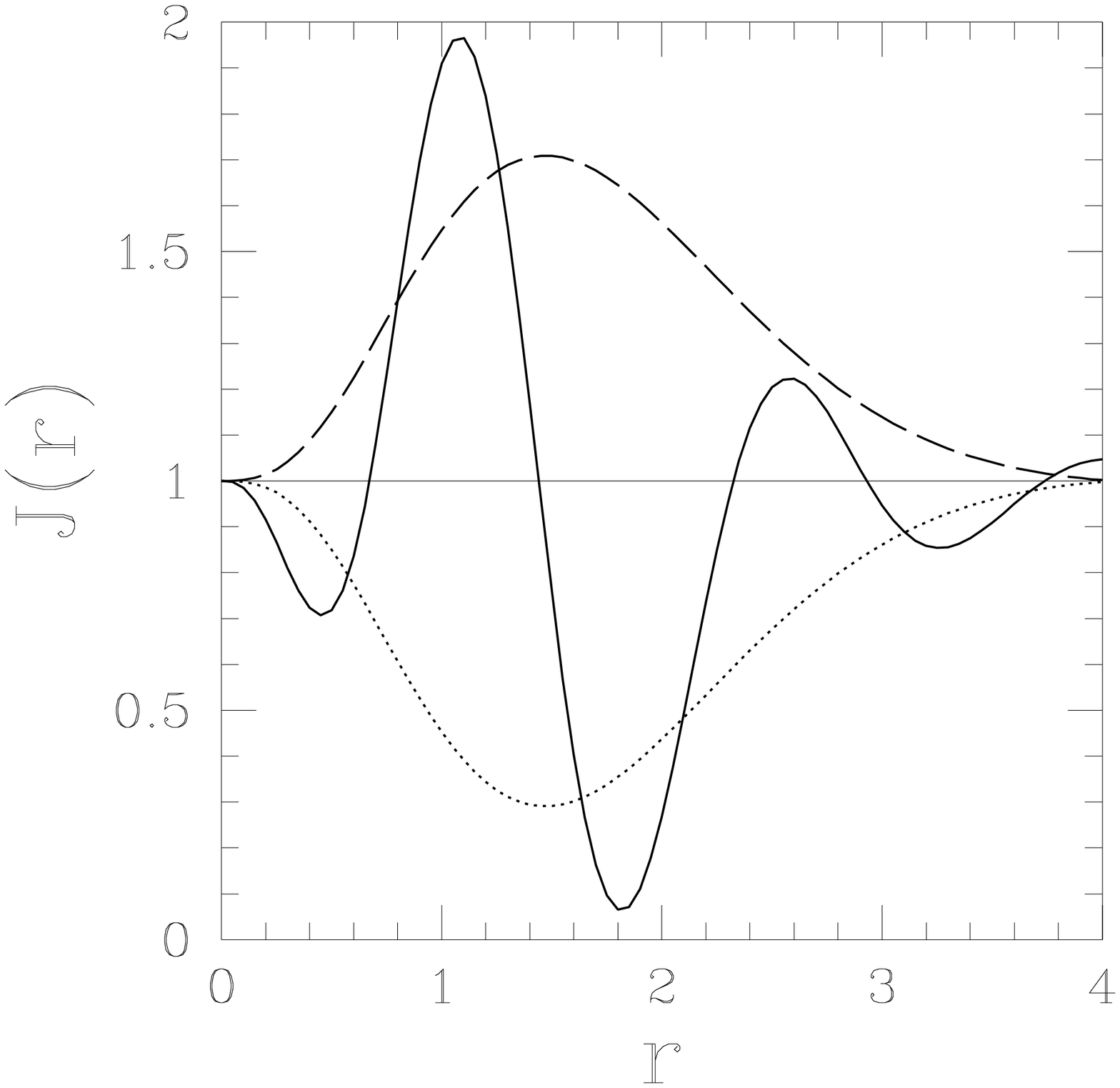} \end{minipage}
\end{center}
\caption[]{\label{fig:gauss} Above we  see the two--point  correlation
function $\xi_2(r)$ of  an anticorrelated point distribution (dashed),
a correlated point  distribution (dotted), and  of a process mimicking
the  cluster distribution (solid); $r$  in  arbirary units. Below  the
corresponding   $J(r)$ with   $\overline{\rho}=1$   in the    Gaussian
approximation are depicted.}
\end{figure}
For  a   stationary  point distribution   $1-F(r)$    is equal to  the
probability that no galaxy is inside a sphere $\CB_r$ with radius $r$:
\begin{equation}
F(r) = 1-\rho_0(\CB_r),
\end{equation}
with $\rho_0(\CB_r)$    being the    void   probability    function
{}\cite{white:hierarchy}.
Similarly,  $1-G(r)$ is  equal to  the probability  that   there is an
object  at $\bx\in\CX$ {\em and} that  there is no other object inside
the sphere $\CB_r(\bx)$ centered on $\bx$:
\begin{equation}
G(r) = 1 - \frac{\rho_1(\bx~|~\CB_r(\bx))}{\overline{\rho}} 
\rho_0(\CB_r(\bx)).
\end{equation}
$\rho_1(\bx~|~\CB_r(\bx))$  is the density  that we observe a point at
$\bx$ under the condition that $\CB_r(\bx)$ is empty\footnote{Assuming
a  stationary point distribution   we  can  choose  $\bx$  to  be  the
origin.}.  Therefore we obtain {}\cite{sharp:holes}:
\begin{equation} \label{eq:Jcorr}
J(r) = \frac{\rho_1(\bx~|~\CB_r(\bx))}{\overline{\rho}}.
\end{equation}
Following   {}\scite{stratonovich:topicsI}   we   can  express     the
conditional density  $\rho_1$ in terms  of the  $n$--point correlation
functions,      the   normed      cumulants,   $\xi_n$   (see     also
{}\pcite{white:hierarchy}):
\begin{eqnarray} \label{eq:rho1}
\lefteqn{\rho_1(\bx~|~\CB_r(\bx)) = \overline{\rho} - 
\sum_{n=1}^\infty \frac{(-\overline{\rho})^{n+1}}{n!} \times} \nonumber\\
& &\times \int_{\CB_r(\bx)}\d^3x_1 \dots \int_{\CB_r(\bx)}\d^3x_n\ 
\xi_{n+1}(\bx,\bx_1,\ldots,\bx_{n})  .
\end{eqnarray}
A Gaussian approximation, i.e.\ $\xi_n=0$ for $n>2$, yields
\begin{equation} \label{eq:Jgauss}
J(r) \approx 1 - \overline{\rho}\ 2\pi\ \int_0^r \d s\ s^2 \xi_2(s).
\end{equation}
In  Fig.~\ref{fig:gauss}   we   show  several  two--point  correlation
functions    $\xi_2(r)$,  satisfying    the normalization    condition
$\int_0^\infty \d s\ s^2 \xi_2(s)=0$, and  the corresponding $J(r)$ in
the  Gaussian approximation.  From  the third example  (solid lines in
Fig.~\ref{fig:gauss}) we    see that a  point   distribution  which is
correlated (clustered)  on small  scales  may show  anticorrelation on
large scales, such that $J(r)<1$ for small and $J(r)>1$ for large $r$.
Two simple examples  of a  correlated distribution with  $\xi_2(r)>0$,
and of an  anticorrelated distribution with  $\xi_2(r)<0$ clearly show
the expected $J(r)<1$ and $J(r)>1$ respectively.

\subsection{Beyond the Gaussian approximation}

\begin{figure}
\begin{center}
\epsfxsize=6cm
\begin{minipage}{\epsfxsize} \epsffile{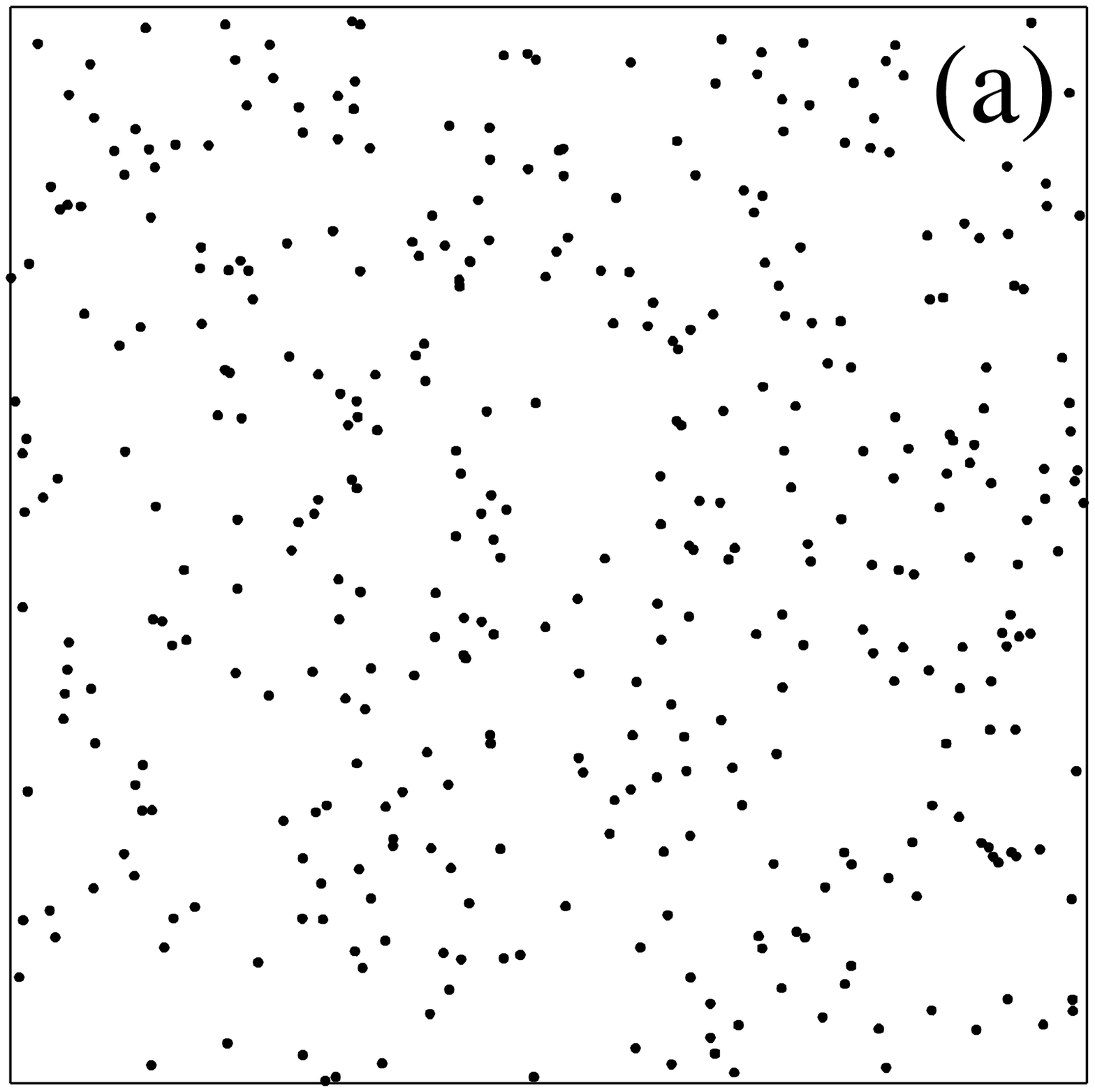} \end{minipage} 
\epsfxsize=6cm
\begin{minipage}{\epsfxsize} \epsffile{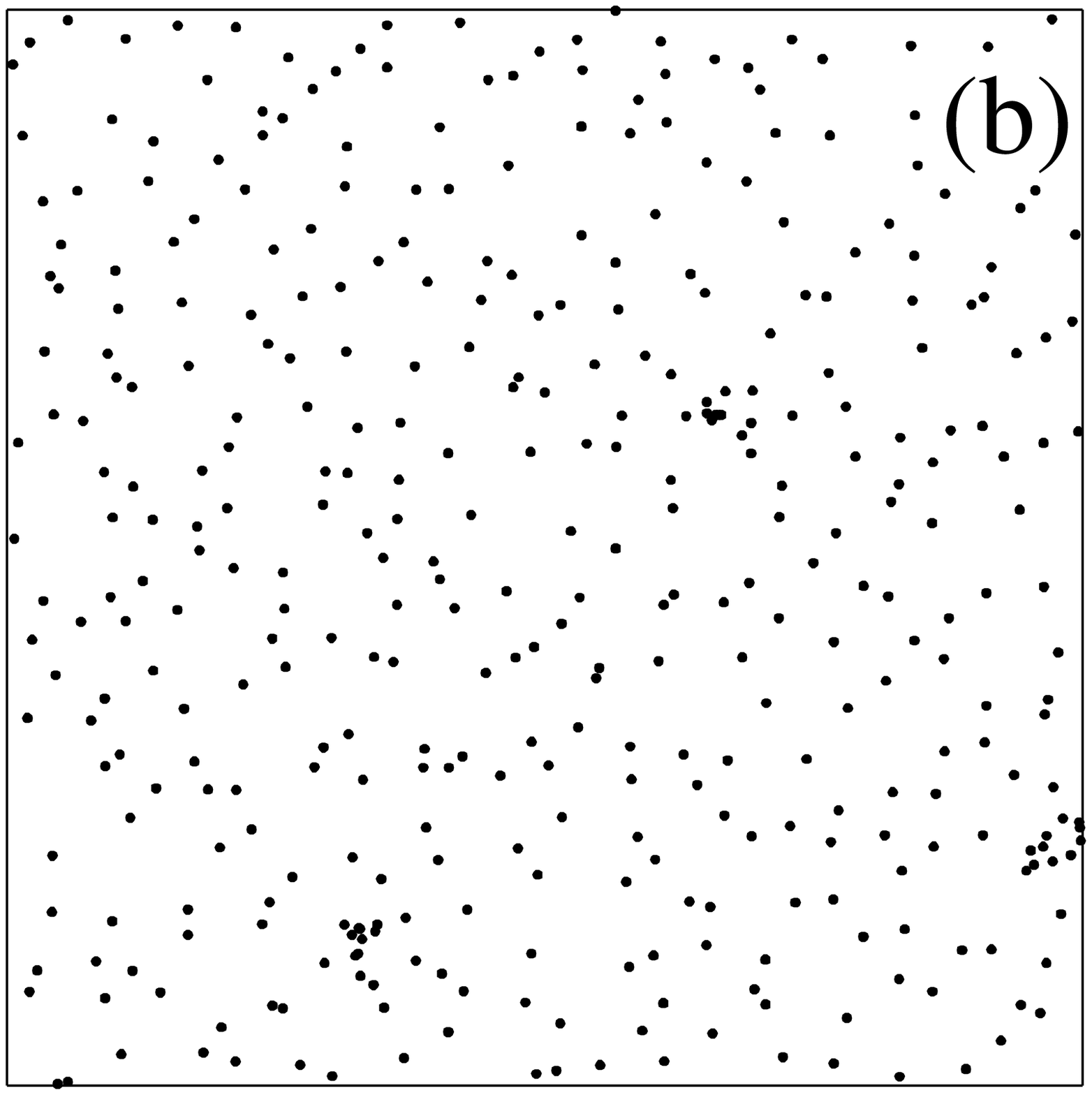} \end{minipage}
\end{center}
\caption[]{\label{fig:2Dvisual}  In    plot  (a)  randomly distributed
points and in plot (b) points from realization  of the regular example
by {}\protect\scite{baddeley:cautionary} are displayed.}
\end{figure}

\begin{figure}
\begin{center}
\epsfxsize=6cm
\begin{minipage}{\epsfxsize} \epsffile{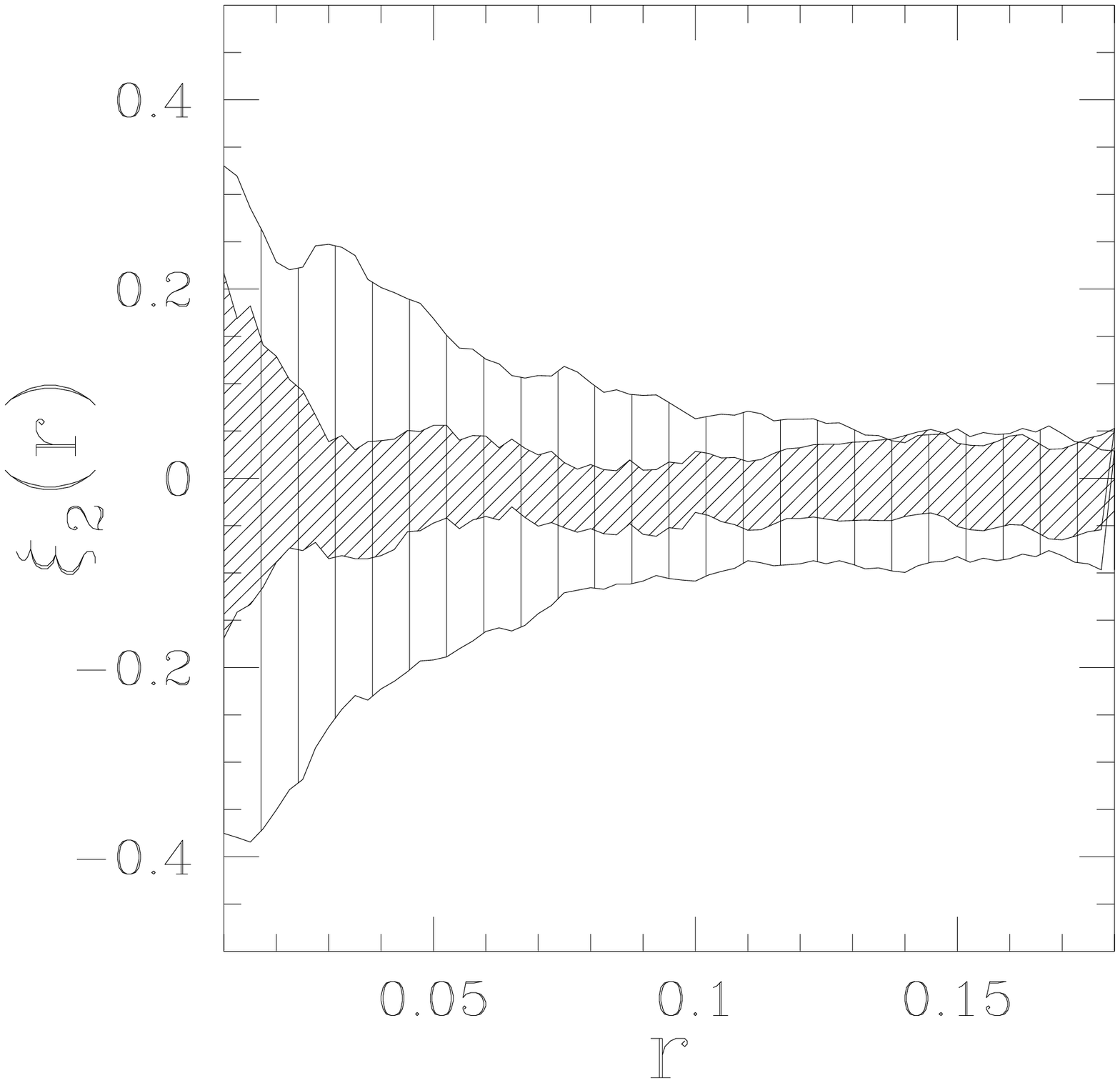} \end{minipage} 
\epsfxsize=6cm
\begin{minipage}{\epsfxsize} \epsffile{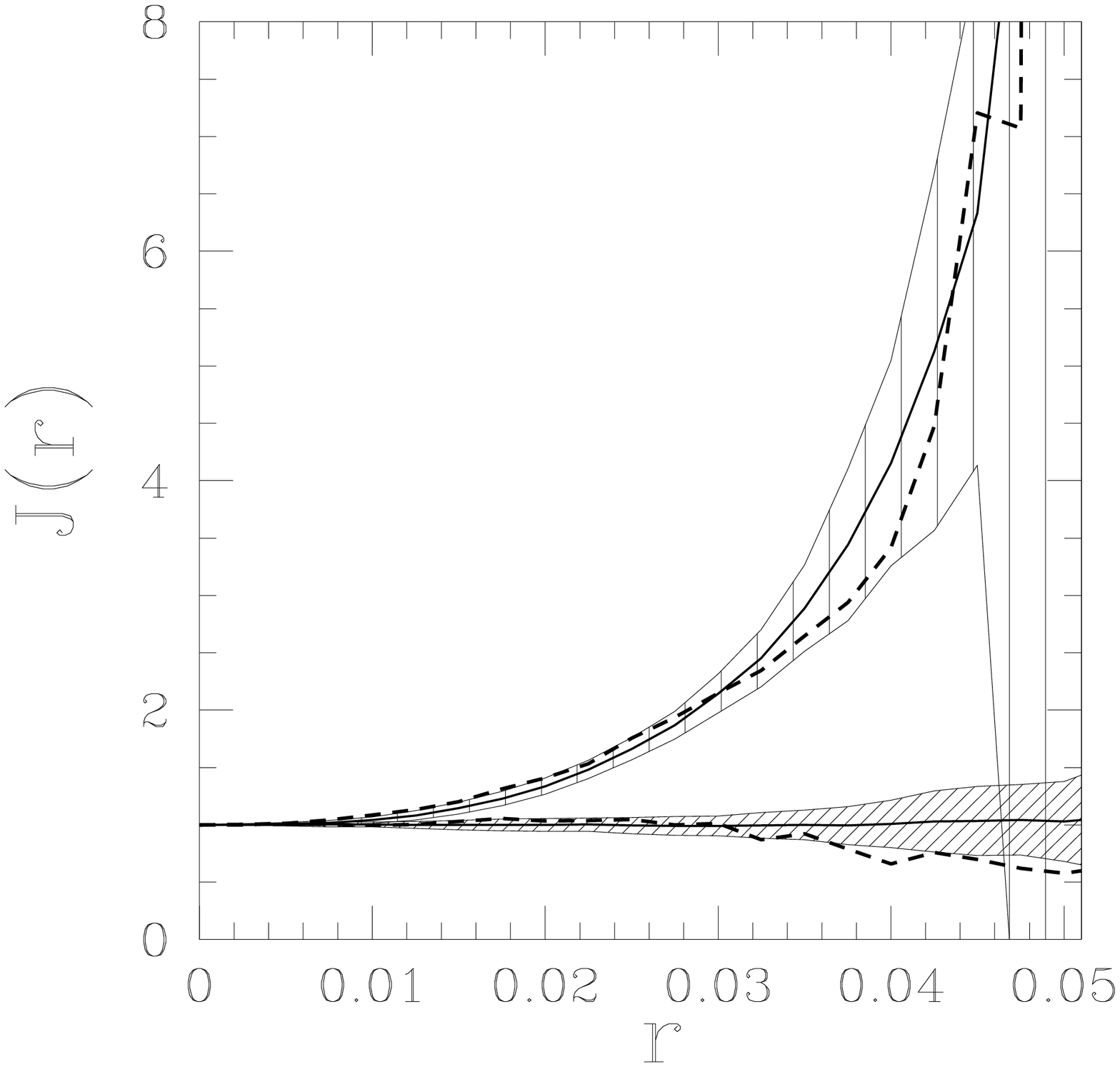} \end{minipage}
\end{center}
\caption[]{   \label{fig:2DxiJ}   The    plots show   the   two--point
correlation  function $\xi_2(r)$ on top,  and $J(r)$ at the bottom for
the random   points (dark shaded)   and  for the  regular  example  by
{}\protect\scite{baddeley:cautionary}   (light     shaded).  The areas
correspond to the  1$\sigma$--error estimated from fifty realizations.
The dashed lines display the $J(r)$ for  the point distributions shown
in Fig.~\ref{fig:2Dvisual}; $r$ is in units of  the side length of the
square.}
\end{figure}

According to Eqs.~(\ref{eq:Jcorr})  and (\ref{eq:rho1}) $J(r)$ depends
on  correlations of    arbitrary   order.   Therefore,  a     Gaussian
approximation to $J(r)$  may be misleading.   We illustrate  this with
two   different    point distributions:   a   Poisson (i.e.\   random)
distribution  of points,    and   points  given  by  the  model     of
{}\scite{baddeley:cautionary}.   Both point  distributions exhibit the
same   two--point    characteristics,     but     the   example     of
{}\scite{baddeley:cautionary} is regular by construction.  To generate
a           realization    of    the     point      distributions   by
{}\scite{baddeley:cautionary}   we    divide  the   unit   square into
$20\times20$ cells and  randomly place 0,  1,  or 10 points into  each
cell, with a probability of $1/10$, $8/9$, and $1/90$ respectively.
In Fig.~\ref{fig:2Dvisual} we  display these  point distributions with
388 points in a square.  By visual inspection the  set of points given
by {}\scite{baddeley:cautionary}  shows  a regular  structure.  Larger
voids are only seen in the Poisson distribution.
We   obtain    $J(r)\ge1$  for    the  regular   point    set, clearly
distinguishable from the $J(r)=1$  for the Poisson  distributed points
(Fig.~\ref{fig:2DxiJ}).   Both have   the  same two point  correlation
function $\xi_2(r)=0$  by construction. No scale  can be  deduced from
the two--point correlation   function      $\xi_2(r)$ as  seen      in
Fig.~\ref{fig:2DxiJ}.  Since the number density and  $\xi_2(r)$ are equal  in
both point sets, the  difference  in $J(r)$ results  from  high--order
correlations only.
$J(r)$ and its  variance diverge near the  intrinsic scale $r=1/20$ of
this  specific  regular point  distribution, since  $F(r)$  approaches
unity.

If we estimate $F(r)$ from one realization of  a point process, $F(r)$
becomes unity  when $r$ becomes larger  than the radius of the biggest
empty sphere which fits inside the sample.  Similarly, we get $G(r)=1$
for $r$ larger than the largest distance between neighbouring objects.
Therefore  the  estimate of the   $J$--function   from a single  point 
set becomes undetermined beyond  these  radii.  Since  $J(r)$ is  a global
measure we are  still  able to detect   regular structures, as  global
features of a point pattern, from the $J(r)\ge 1$  for radii $r$ below
the intrinsic scale, here $1/20$.

As  discussed    in    {}\scite{bedford:remark}, $J(r)=1$     does not
necessarily imply Poisson distributed points.
The     morphological     measure       $J(r)$   was      used      by
{}\scite{kerscher:fluctuations}    to   investigate      large   scale
fluctuations  in     the  galaxy   distribution   and    earlier    by
{}\scite{sharp:holes}   to  test   a  hierarchical model    for galaxy
clustering.

Since all real  astronomical catalogues are  spatially limited we have
to    use     edge--corrected     estimators     as     detailed    in
{}\scite{kerscher:fluctuations}.     The   rationale   behind    these
estimators is to use only the points whose possible nearest neighbours
are contained in the sample window.   With these estimators we neither
make any assumptions about  the exterior of our  sample, nor do we use
weighting  schemes.  However, our analysis  of the supercluster sample
is restricted to a radial distance of at most 60--70\hMpc.

\section{Results}
\label{sect:results}

\subsection{The supercluster sample}

We       use         the   supercluster       sample     compiled   by
{}\scite{einasto:supercluster_data} with      220 superclusters out to
$z=0.12$.   The   sample  was  generated   with  a friend--of--friends
algorithm using a  linking length of 24\hMpc\  from an earlier version
of the cluster sample by  {}\scite{andernach:current} and includes all
superclusters of at least two member  clusters.  A detailed discussion
of  the sample  is given   in {}\scite{einasto:supercluster_data}.  We
limit our analysis to a region within galactic latitude $|b|>20^\circ$
and a  maximum radial distance of  330 \hMpc. As directly suggested by
the sample  geometry we   perform   our analysis  separately  for  the
northern and    southern   parts   (in galactic   coordinates).     95
superclusters  enter into the northern part  (mainly Abell sample) and
116 superclusters into the southern part (mainly ACO sample).
The selection   effects are  modeled  with an  independent  radial and
angular selection function (see {}\pcite{einasto:supercluster_II}).

\subsection{Regular structures?} 
\label{sect:results-regular}

\begin{figure}
\begin{center}
\epsfxsize=8cm
\begin{minipage}{\epsfxsize} \epsffile{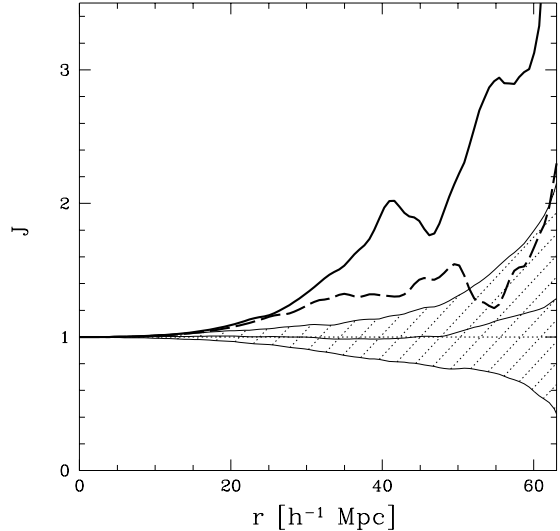} \end{minipage}
\end{center}
\caption[]{\label{fig:sc_J}   $J(r)$   for the northern  part
(dashed),  the  southern  part (solid),  and   a Poisson  sample (shaded
area) are displayed. The curves  are smoothed with  a triangular kernel 
with a total width of 3\hMpc.}
\end{figure}

In  Fig.~\ref{fig:sc_J}   the   values of $J(r)$    for  the supercluster
distribution  are  plotted      together   with the     average    and
$1\sigma$--error  of 99 realizations of  a (pure) Poisson distribution
with the same sample size and geometry.
A $J(r)$ larger than unity, as expected  for a regular distribution of
the points  (see Sect.~\ref{sect:methods}), is   clearly seen.  $J(r)$
for both parts is above one, lying outside the $1\sigma$--range of the
Poisson distribution on scales from $15\hMpc$ to $50 \hMpc$.  The kink
in $J(r)$  at $r=45\hMpc$ in the southern  part and at 55\hMpc\ in the
northern part  indicates    the typical scale on which  the    nearest
supercluster is situated. This agrees  with the median distance to the
nearest   poor     supercluster   of  $45\hMpc$    as    estimated  by
{}\scite{einasto:supercluster_data}.
As discussed in  Sect.~\ref{sect:methods} $J(r)$ becomes unreliable on
scales beyond $60 \hMpc$ (see also Fig.~\ref{fig:sc_J}).
With   a   (nonparametric)  Monte--Carlo  test,     as  described   by
{}\scite{besag:simple},  we   conclude with 95\%  confidence  that the
superclusters  given by  {}\scite{einasto:supercluster_data} are   not
compatible with  Poisson   distributed points  with  the same   number
density.
We will  see that this is  {\em not} decisive, since  up to now we did
not include selection and construction effects.

\begin{figure}
\begin{center}
\epsfxsize=8cm
\begin{minipage}{\epsfxsize} \epsffile{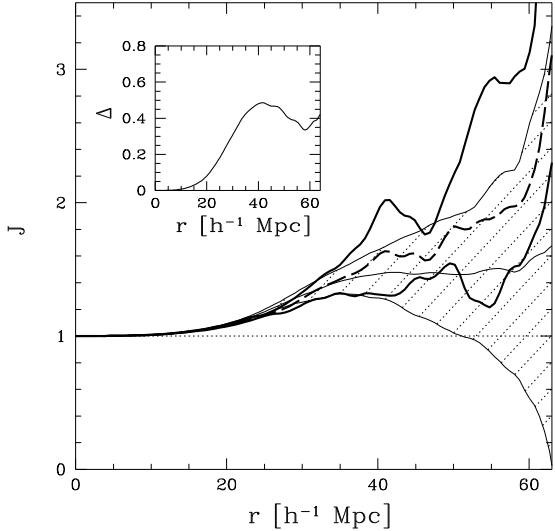} 
\end{minipage}
\end{center}
\caption[]{\label{fig:sc_max} The lower solid  line is the  result for
the  northern part, the  upper solid line  the result for the southern
part and the dashed  line the result for the  whole sample. The shaded
area  marks the $1\sigma$--range estimated  from  99 mock supercluster
realizations. The inset plot shows the difference $\Delta$ between the
mean  values of $J(r)$  for  the pure Poisson    process and the  mock
superclusters samples.  All   curves are  smoothed  with  a triangular
kernel with a total width of 3\hMpc.}
\end{figure}

To  test the influence of the  selection effects  and the construction
process  we  generate 99  ``mock supercluster  catalogues''.  We start
with  Poisson distributed   points  within  a   sphere   of 370\hMpc\
incorporating  the radial and  angular selection effects of the galaxy
cluster catalogue given  by {}\scite{einasto:supercluster_II}; then we
apply a friend--of--friends procedure  with linking length of 24\hMpc\
to identify the mock ``superclusters''.
As seen   in   Fig.~\ref{fig:sc_max},   using a    friend--of--friends
algorithm, we generate   an  empty sphere  with  radius   of  at least
24\hMpc\ around  each supercluster  center, introducing  an artificial
anticorrelation, leading to $J(r) > 1$.
Therefore,  the  regularity   seen  in  the   northern  part  of   the
supercluster sample up  to scales of 60\hMpc\  is  at least partly  an
artifact of the construction.  Still the southern  part shows a $J(r)$
above the mean  $J(r)$ of the  mock superclusters, mostly  outside the
$1\sigma$--range, but a definite statement with a significance of 95\%
(roughly $2\sigma$) is no longer possible.

\section{Discussion and Conclusion}
\label{sect:conclusion}

We have  shown that the   statistical properties of  the  supercluster
distribution as   given by  {}\scite{einasto:supercluster_data}    are
seriously  affected  by the  construction  with  a friend--of--friends
procedure.    This  is not  astonishing  since   the linking length of
24\hMpc\  is already one  fifth  of the   claimed regularity scale  of
120\hMpc.
The distinction between regular  and clustered point patterns with the
$J$--function is unambiguous  for  a homogeneous and   isotropic point
distribution. In such a case the borderline is given by $J(r)=1$.
Our  procedure for  generating  mock supercluster samples,  where  we
start with  a Poisson sample,  include the selection effects, and redo
the supercluster  identification  with a friend--of--friends algorithm
in   the same way   as  for the real  cluster   sample,  results in  a
$J(r)\ne1$ even  though   we  started  from a  Poisson   distribution.
Although   it  is  plausible,  that   such  mock--supercluster samples
describe  the   borderline between clustered  and  regular structures,
there is no proof of this assertion.
The apparent regularity  in  the northern  part of  the sample can  be
explained as a result    of these construction effects,   whereas  the
southern  part still shows  a trend towards regular structures outside
the $1\sigma$--range.

The  results for the   oscillating two--point correlation function  of
galaxy clusters and, correspondingly the   peak in the power  spectrum
were   obtained with estimators using   weighting schemes and boundary
corrections, which rely heavily on the assumption  of homogeneity.  Up
to  now  there   is   no  reliable    way to prove    this   from  the
three--dimensional distribution of galaxies  and clusters.   There are
some  hints  that the   universe  reaches homogeneity on  scales above
several  hundreds       of~\hMpc\   (see    the      discussions    by
{}\pcite{guzzo:homogeneous}  versus     {}\pcite{labini:scale}).    We
adopted a conservative point of view  and used estimators which do not
make any assumptions about the  distribution of superclusters  outside
the sample window.
In  Sect.~\ref{sect:methods} we showed that  the  $J$--function can be
estimated from one  point set only for scales  smaller than the radius
of the  largest   void.   Therefore, we   do not    reach  the claimed
regularity  scale at  120\hMpc. Still, the   measure  $J(r)$ gives  us
information about  global properties of the supercluster distribution,
in our case a tendency towards regular structures.

We analyzed the  distribution of clusters of  the more recent redshift
compilation by {}\scite{andernach:current} with the $J$--function.  We
found   the  expected clumping of   galaxy   clusters, as indicated by
$J(r)\le1$.  Qualitatively, the   $J(r)\le1$ may be   explained with a
$\xi_2(r)>0$  and the Gaussian approximation in Eq.~(\ref{eq:Jgauss}).
This  clumping  out to scales  of  40\hMpc\ is  confined mainly to the
interior of the superclusters. Isolated "field" clusters
were  not included in the  supercluster  sample but may contribute to
the correlation seen up to  scales of 50\hMpc in the cluster samples.  
One hierarchical level
higher, the supercluster  centers  themselves show a tendency  towards
regular structures.  Again   this can be explained  qualitatively with
the Gaussian approximation (see Fig.~\ref{fig:gauss}).
A theoretical example  illustrating such  a  hierarchical property  is
given by Neyman--Scott  processes {}\cite{neyman:statistical}: In such
a process the overall distribution  of points shows correlation (i.e.\
$\xi(r)>0$ for small $r$), but the cluster centers of these points are
distributed randomly by construction.

Unlike the  two--point correlation function the $J$--function
incorporates information  stemming from high  order correlations.  Our
example in  Fig.~\ref{fig:2DxiJ} illustrates, that a regular structure
detected unambigously with the $J$--function  may not be visible in an
analysis with the two--point correlation function $\xi_2(r)$ alone.

Another problem is the  fluctuations between the northern and southern
parts of the sample. This may be attributed to the different selection
effects entering the  Abell and ACO  parts of the sample, probably due
to the  different sensitivity of the  photo plates used.   However, in
the case of  the IRAS 1.2~Jy  galaxy catalogue such  fluctuations were
shown   to        be      real    on     scales        of    200\hMpc\
{}\cite{kerscher:fluctuations}.
Also, {}\scite{zucca:esp-ii} find from  the ESP survey, that  at least
in the southern hemisphere the local density is  below the mean sample
density out to 140\hMpc.
If we assume that the fluctuations  decrease on scales above 200\hMpc,
the   finding of regular structures  on  such large  scales is a great
challenge   to   the standard scenarios   of    structure formation by
gravitational  instability,   starting from Gaussian  initial  density
fluctuations.    Implications  of these    regular structures  for the
standard scenarii  of     structure  formation  are     discussed   in
{}\scite{einasto:supercluster_III} and {}\scite{szalay:walls}.

\section*{Acknowledgement}
I want to thank H.~Andernach for suggestions on the text, C.~Beisbart,
T.~Buchert, M.~Einasto, V.J.~Mart\'{\i}nez,  M.J.~Pons--Border\'{\i}a,
R.~Trasarti--Battistoni,  and especially J.~Schmalzing, H.~Wagner  and
the referee for  valuable  comments.  H.~Andernach and E.~Tago  kindly
provided a  suitable  extraction from the   Dec.~1997 version of their
Abell/ACO redshift  compilation.     I acknowledge support from    the
Sonderforschungsbereich    SFB  375  f\"ur    Astroteilchenphysik  der
Deutschen   Forschungsgemeinschaft  and   by  the  Acci\'on  Integrada
Hispano--Alemana HA-188A (MEC).


\begin{thebibliography}{{{Sylos Labini} \bgroup et al.\egroup }{1998}}

\bibitem[\protect\citefmt{Andernach \& Tago}{1998}]{andernach:current}
Andernach H., Tago E.: 1998, In: \emph{Proc. Large Scale Structure: Tracks
  and Traces, Potsdam, Germany} (Singapore), M\"uller V., Gottl\"ober S.,
  M\"ucket J.~P., Wambsgans J. (eds.), World Scientific, in press,
  astro-ph/9710265

\bibitem[\protect\citefmt{Baddeley \& Silverman}{1984}]{baddeley:cautionary}
Baddeley A.~J., Silverman B.~W., 1984, Biometrics 40, 1089

\bibitem[\protect\citefmt{Bedford \& {van den Berg}}{1997}]{bedford:remark}
Bedford T., {van den Berg} J., 1997, Adv.\ Appl.\ Prob. 29, 19

\bibitem[\protect\citefmt{Besag \& Diggle}{1977}]{besag:simple}
Besag, J., Diggle, P.~J., 1977, Appl.\ Statist. 26, 327

\bibitem[\protect\citefmt{Broadhurst \bgroup et al.\egroup
  }{1990}]{broadhurst:large-scale}
Broadhurst T.~J., Ellis R.~S., Koo D.~C., Szalay A.~S., 1990,
  \providecommand{\nat}{Nature}{\nat} 343, 726

\bibitem[\protect\citefmt{Einasto \bgroup et al.\egroup
  }{1997a}]{einasto:supercluster_II}
Einasto J., Einasto M., Frisch P. et~al., 1997a,
  \providecommand{\mnras}{Mon.\ Not.\ Roy.\ Astron.\ Soc.}{\mnras} 289,
  801

\bibitem[\protect\citefmt{Einasto \bgroup et al.\egroup
  }{1997b}]{einasto:supercluster_III}
Einasto J., Einasto M., Frisch P. et~al., 1997b,
  \providecommand{\mnras}{Mon.\ Not.\ Roy.\ Astron.\ Soc.}{\mnras} 289,
  813

\bibitem[\protect\citefmt{Einasto \bgroup et al.\egroup
  }{1997c}]{einasto:120mpc}
Einasto J., Einasto M., Gottl{\"o}ber S. et~al., 1997c,
  \providecommand{\nat}{Nature}{\nat} 385, 139

\bibitem[\protect\citefmt{Einasto \bgroup et al.\egroup
  }{1997d}]{einasto:supercluster_data}
Einasto M., Tago E., Jaaniste J. et~al., 1997d,
  \providecommand{\aas}{Astron.\ Astrophys.\ Suppl.}{\aas} 123, 119

\bibitem[\protect\citefmt{Fetisova \bgroup et al.\egroup
  }{1993}]{fetisova:features}
Fetisova T.~S., {\kern0exYu}.~Kuznetsov D., Lipovetskii V.~A. et~al., 1993,
  Astron.\ Lett. 19(3), 198

\bibitem[\protect\citefmt{Guzzo}{1997}]{guzzo:homogeneous}
Guzzo L., 1997, New Astronomy 2(6), 517

\bibitem[\protect\citefmt{Hansen \& McDonnald}{1986}]{hansen:theory}
Hansen J.~P., McDonnald I.~R., 1986, \emph{Theory of simple liquids},
  Academic Press, New York and London

\bibitem[\protect\citefmt{Kerscher \bgroup et al.\egroup
  }{1997}]{kerscher:fluctuations}
Kerscher M., Schmalzing J., Buchert T., Wagner H. 1997,
  \providecommand{\aanda}{Astron.\ Astrophys.}{\aanda} in press,
  astro-ph/9704028

\bibitem[\protect\citefmt{Kopylov \bgroup et al.\egroup
  }{1988}]{kopylov:possible}
Kopylov A.~I., {\kern0exYu}.~Kuznetsov D., Fetisova T.~S., Shvartsman
  V.~F.: 1988, In: \emph{Large Scale Structure of the Universe}, 
  Audouze J.~A. et~al. (ed.), IAU, pp.~129

\bibitem[\protect\citefmt{Mecke \bgroup et al.\egroup }{1994}]{mecke:robust}
Mecke K.~R., Buchert T., Wagner H., 1994, \providecommand{\aanda}{Astron.\
  Astrophys.}{\aanda} 288, 697

\bibitem[\protect\citefmt{Mo \bgroup et al.\egroup }{1992}]{mo:typical_scales}
Mo H.~J., Deng Z.~G., Xia X.~Y. et~al., 1992,
  \providecommand{\aanda}{Astron.\ Astrophys.}{\aanda} 257, 1

\bibitem[\protect\citefmt{Neyman \& Scott}{1958}]{neyman:statistical}
Neyman J., Scott E.~L., 1958, J.\ R.\ Stat.\ Soc. 20, 1

\bibitem[\protect\citefmt{Sharp}{1981}]{sharp:holes}
Sharp N., 1981, \providecommand{\mnras}{Mon.\ Not.\ Roy.\ Astron.\
  Soc.}{\mnras} 195, 857

\bibitem[\protect\citefmt{Stratonovich}{1963}]{stratonovich:topicsI}
Stratonovich R.~L., 1963, \emph{Topics in the theory of random noise} Vol.~1,
  Gordon and Breach, New York

\bibitem[\protect\citefmt{{Sylos Labini} \bgroup et al.\egroup
  }{1998}]{labini:scale}
{Sylos Labini} F., Montuori M., Pietronero L., 1998, Physics Rep. 293,
  61

\bibitem[\protect\citefmt{Szalay}{1997}]{szalay:walls}
Szalay A.~S.: 1997, In: \emph{Proc. of the 18th Texas Symposium on
  Relativistic Astrophysics} (New York), Olinto, A. (ed.), AIP

\bibitem[\protect\citefmt{{van Lieshout} \& Baddeley}{1996}]{vanlieshout:j}
{van Lieshout} M.~N.~M., Baddeley A.~J., 1996, Statist.\ Neerlandica 50,
  344

\bibitem[\protect\citefmt{White}{1979}]{white:hierarchy}
White S. D.~M., 1979, \providecommand{\mnras}{Mon.\ Not.\ Roy.\ Astron.\
  Soc.}{\mnras} 186, 145

\bibitem[\protect\citefmt{Zucca \bgroup et al.\egroup }{1997}]{zucca:esp-ii}
Zucca E., Zamorani G., Vettolani G. et~al., 1997,
  \providecommand{\aanda}{Astron.\ Astrophys.}{\aanda} 326, 477

\end{thebibliography}

\providecommand{\bysame}{\leavevmode\hbox to3em{\hrulefill}\thinspace}

\end{document}